\documentclass[twoside,11pt]{article}

\usepackage{ulem} 

\usepackage{jmlr2e}

\ShortHeadings{Combined parameter and state inference with ABC}{Ebert, Pudlo, Mengersen, Wu and Drovandi}
\firstpageno{1}

\usepackage[utf8]{inputenc}
\usepackage[english]{babel}
\usepackage{graphicx}
\usepackage{amsfonts,amssymb,amsmath,amscd,latexsym}
\usepackage{mathrsfs}
\usepackage{algorithm,algpseudocode}
\usepackage{enumerate}
\usepackage{txfonts}

\newtheorem{theo}{Theorem}

\newtheorem{pro}[theo]{Proposition}

\newcommand{\esp}{\mathbb{E}}
\renewcommand{\epsilon}{\varepsilon}
\newcommand{\ds}{\displaystyle}

\begin{document}

\title{Combined parameter and state inference with automatically calibrated approximate Bayesian computation}

\author{\name Anthony\ Ebert \email anthonyebert@gmail.com \\
       \addr School of Mathematical Sciences\\
       Queensland University of Technology\\
       Brisbane, QLD 4000, Australia
       \AND
       \name Pierre\ Pudlo \email pierre.pudlo@univ-amu.fr \\
       \addr Institut de Mathématiques de Marseille\\
       Aix-Marseille Université, CNRS, Centrale Marseille \\
       Marseille, 13453 Cedex 13, France
       \AND
       \name Kerrie\ Mengersen \email k.mengersen@qut.edu.au \\
       \addr School of Mathematical Sciences\\
       Queensland University of Technology\\
       Brisbane, QLD 4000, Australia
       \AND
       \name Paul\ Wu \email p.wu@qut.edu.au \\
       \addr School of Mathematical Sciences\\
       Queensland University of Technology\\
       Brisbane, QLD 4000, Australia
       \AND
       \name Christopher\ Drovandi \email c.drovandi@qut.edu.au\\
       \addr School of Mathematical Sciences\\
       Queensland University of Technology\\
       Brisbane, QLD 4000, Australia
     }

\editor{}

\maketitle

\begin{abstract}
  State space models contain time-indexed parameters, termed states, as well as static parameters, simply termed parameters. The problem of inferring both static parameters as well as states simultaneously, based on time-indexed observations, is the subject of much recent literature. This problem is compounded once we consider models with intractable likelihoods. In these situations, some emerging approaches have incorporated existing likelihood-free techniques for static parameters, such as approximate Bayesian computation (ABC) into likelihood-based algorithms for combined inference of parameters and states. These emerging approaches currently require extensive manual calibration of a time-indexed tuning parameter: the acceptance threshold.
  
  We design an SMC$^2$ algorithm (Chopin et al., 2013, JRSS B) for likelihood-free approximation with automatically tuned thresholds. We prove consistency of the algorithm and discuss the proposed calibration. We demonstrate this algorithm's performance with three examples. We begin with two examples of state space models. The first example is a toy example, with an emission distribution that is a skew normal distribution.  The second example is a stochastic volatility model involving an intractable stable distribution. The last example is the most challenging; it deals with an inhomogeneous Hawkes process.
\end{abstract}






\section{Introduction}


\subsection{Approximate Bayesian computation (ABC)}
\label{sub:ABC}

Approximate Bayesian computation \citep{tavare_inferring_1997,marin2012approximate,sisson2018handbook} is a Monte Carlo method for performing parameter inference where the likelihood $p(y|\theta)$ is intractable. Suppose we have some latent process $x$, then Bayesian inference becomes more difficult when the multidimensional integral
\[
  p(y|\theta) = \int p(x,y|\theta) \text{d}x
\]
cannot be computed explicitly. The output of ABC algorithms is a sample from an approximation of the posterior distribution $\pi(\theta|y)$. This approximation relies on parameter samples $\theta$ drawn from the prior, and samples $(\tilde x, \tilde y) | \theta$ of the latent process and the observed process drawn from the stochastic model given $\theta$. Given these simulations, ABC algorithms derive approximations of the posterior $\pi(\theta|y)$ and of other functionals of the posterior such as moments or quantiles. The most profound shortcoming with ABC is that approximating the conditional distribution of $\theta$ given $\tilde y = y$ requires that at least some simulations $\tilde y$ fall into the neighborhood of $y$. If the data space is of high dimension, we face the curse of dimensionality, namely that it is almost impossible to get a simulated dataset $\tilde y$ near the observed $y$. To circumvent this problem, ABC schemes perform a non-invertible mapping from the observed and simulated datasets onto a space of lower dimension via some summary statistics $s:=\eta(y)$ and $\tilde s:=\eta(\tilde y)$, and set a metric $d(s, \tilde s)$ on this space. Thus, ABC cannot recover anything better than $\pi_0(\theta|y):=\pi(\theta|\tilde s = s)$. For instance, the accept-reject ABC algorithm samples the distribution given the event $A_\epsilon = \{d(\tilde s, s)\le \epsilon\}$, i.e., $\pi_\epsilon(\theta|y):=\pi(\theta|d(\tilde s,s)\le \epsilon)$, where $\epsilon$ is a threshold that has to be set by the user. This distribution tends to $\pi_0(\theta|y)$ when $\epsilon\to0$. But, if we want a sample of size $N$ from $\pi_\epsilon(\theta|y)$, the accept-reject ABC algorithm requires on average a number of simulations from the prior given by $N_\text{prior}:=N/\pi(A_\epsilon)$. Since $\pi(A_\epsilon)$ can decrease very quickly towards $0$ when $\epsilon\to0$, the ideal value $\epsilon\approx 0$ may be difficult to achieve with the accept-reject ABC algorithm. In practice, $\epsilon$ corresponds to a quantile of the distances between the observed $s$, the simulated $\tilde s$: $N_\text{prior}$ is set by the computational budget, and $N$ is the desired number of posterior samples. The threshold is set so that $\pi(A_\epsilon)= N/N_\text{prior}$, which means that $\epsilon$ is the quantile of order $N/N_\text{prior}$ of the distance to the observed dataset.

Various sampling algorithms that also target $\pi_\epsilon(\theta|y)$ have been proposed in the literature to improve the accept-reject algorithm, particularly where the parameter space is high dimensional. The ABC-MCMC \citep{marjoram_markov_2003} is rather popular, but it can perform poorly if the threshold $\epsilon$ is small. Sequential Monte Carlo algorithms have been proposed as an alternative to these methods: \citet{sisson_sequential_2007}, \citet{beaumont2009adaptive}, \citet{drovandi_estimation_2011}, \citet{del2012adaptive}. These sequential algorithms decrease the threshold gradually and tune the proposal distribution for the parameter $\theta$ to get closer and closer to the approximate posterior $\pi_0(\theta|y)$. Yet these algorithms suffer from two problems:
\textit{(i)} at each iteration, they have to draw many complete datasets $\tilde y$, which can be heavy in terms of computation time; \textit{(ii)} they still require a summary function $\eta$ that projects the complete datasets into a space of much smaller dimension. With this projection, therefore, comes a possible loss of information, since we rarely have a theoretical guarantee that the statistic is sufficient.

State space models comprise one class of statistical inference problem where the parameter space is very high dimensional. There are Monte Carlo methods for Bayesian analysis of state space models, which we will explore in the next section. These methods usually target the distribution $p(x_t,\theta|A_{\epsilon_{1:t}})$ sequentially. The posterior distribution can then be recovered by trivial marginalisation of the final target. However, the thresholds $\epsilon_1,\ldots, \epsilon_T$ are much more difficult to tune in this setting and may require multiple pilot runs before achieving acceptable results. 

\subsection{Combined parameter and state inference}

In a typical state space model, two types of parameters coexist: one type varies on a time index (states), and the other type is independent of time (fixed static parameters, hereafter simply termed parameters). Observations are available on the same indexed time domain as the state. Common inference techniques perform inference on the state, conditional on known static parameters. 

State inference is termed smoothing, filtering, or prediction depending on whether the state inference is earlier, cotemporal to, or later than the last observation respectively \citep{sarkka_bayesian_2013}. Usually, interest lies in filtering, state inference conditional on all data processed thus far. The Kalman filter \citep{kalman1960new} is a well-known analytic solution for filtering, but there are strict assumptions on $p(y_t|x_t,\theta)$, which is assumed to be Gaussian in nature. More complex cases typically require a Monte Carlo based approach, termed a particle filter (PF), where a vector of proposed states propagates along with the time index \citep{kitagawa1987non,gordon1993novel}. For recent overviews of the subject see \citet{fearnhead2018particle} and \citet{naesseth2019elements}. 

Unknown static parameters add a surprising degree of complexity to the inference problem. Algorithms for combined parameter and state (CPS) inference have received less attention than filtering, see \citet{kantas2015particle,liu2001combined} for comprehensive overviews. A straightforward approach introduces an artificial non-degenerate transition density for the static parameters and proceeds with the particle filter as though these static parameters were states, along with the genuine state parameters. This approach is problematic as the static parameter samples are no longer guaranteed to be from the true posterior. Furthermore, if the true parameter value is not within the initial draw of parameter proposals then the final weighted set of proposals will not contain the true value. This problem is not evident for the state proposals as they exhibit stochasticity naturally. 

More recent techniques embed a PF for state parameters within a larger algorithm for static parameters.  Examples include particle Markov chain Monte Carlo (PMCMC) \citep{andrieu2010particle,drovandi2016exact}, sequential Monte Carlo (SMC, SMC$^2$) \citep{liu2008monte,del2012adaptive,chopin2013smc2,drovandi2016alive}, and Bayesian indirect inference \citep{martin2019auxiliary}. CPS inference techniques have many applications including ecology \citep{fasiolo2016comparison}, agent-based models \citep{lux2018estimation}, genetic networks \citep{marino2017comparison}, and hydrological models \citep{fenicia2018signature}. 

\subsection{Our contributions and plan}

In this paper we develop an ABC sequential method that automatically tunes the time-indexed thresholds at each iteration of the algorithm.  Section~\ref{sec:Background} provides some theoretical background for state space models, particle filtering and Bayesian static parameter inference within this framework.

The methodology we propose is given in Section~\ref{sec:ABCSMC2_methodology}, which is based on the efficient SMC$^2$ algorithm proposed by \citet{chopin2013smc2}. Our proposal includes the filtering algorithm of \cite{jasra2012filtering} to each particle of a larger SMC that takes into account the uncertainty about static parameters of the state space model. Our proposal is thus an adaptation of the SMC$^2$ algorithm to approximate Bayesian computation. We take care of tuning automatically the time-index thresholds that are required by the algorithm. We investigate some theoretical properties of the method at the end of Section~\ref{sec:ABCSMC2_methodology}, and discuss the effect of the automatic calibration.

Our approach is illustrated on three different examples in Section~\ref{sec:ABCSMC2_examples}: a state space model with skew normal distributions, a stochastic volatility model from the econometric literature, and a Hawkes process. 


\section{Background} \label{sec:Background}

\subsection{State space model} \label{sec:ABCSMC2_background}

The first example of a state space model was a hidden Markov models in the seminal paper by \citet{baum1966statistical}. It is customary to use `state space' notation for state parameters where $x_t$ refers to the value of the state parameter at time $t$. We consider state space models where state transitions are Markovian, conditional on a static parameter $\theta$:
\begin{align*}
p(x_{t+1} | x_{1:t}, \theta) &= p(x_{t+1} | x_t, \theta),
\end{align*}
where the notation $x_{1:t}$ denotes the subset of $x$ from time 1 to time $t$. The state process $x = (x_1, x_2, \dots, x_T)$ is latent, but there is an observed process $y = (y_1, y_2, \dots, y_T)$, also conditioned on $\theta$:
\begin{align}
p(y_{t} | y_{1:t-1}, x_{1:t}, \theta) &= p(y_t | x_t, \theta),
\end{align}
termed the emission distribution. We can see, given $\theta$ and $x_t$, that $y_t$ is independent of its history $y_{1:t-1}$.

In nonlinear and non-Gaussian state space models, none of the distributions given below are explicitly available:
\textit{(i)} the filtering distribution: $p(x_t|y_{1:t}, \theta)$, \textit{(ii)} the smoothing distributions: $p(x_{t-1,t} | y_{1:T}, \theta)$, with $t<T$, and $p(x_{1:T}|y_{1:T},\theta)$,
moreover the likelihood $p(y_{1:T}|\theta)$ is intractable.
We briefly introduce particle filters in Section~\ref{sub:PF} and consider combined parameter and state space inference with the SMC$^2$ in Section~\ref{sub:SMC2}. Likelihood-free versions of these algorithms are discussed in Section~\ref{sub:ABCSSM}.

\subsection{Particle filters} \label{sub:PF}

\begin{algorithm}
	\caption{Sequential importance resampling filter \citep{doucet_sequential_2000} \label{algo:PF}}
		\begin{algorithmic}[1]
	\Require $\theta$; and observations $y_{1:T}$
	\Ensure state samples $(x_t^1,\dots, x_t^{N_x})$ with associated normalised weights $(W_t^1, \dots, W_t^{N_x})$, and marginalised likelihood estimate $\hat p(y_t|y_{1:t-1}, \theta)$
		\For{$n=1,\ldots, N_x$}
			\State draw $x_1^n \sim q_1(\cdot| \theta)$
  			\State set $w_1^n = p(x_1^n|\theta) p(y_1|x_1^n,\theta) \Big/ q_1(x_1^n| \theta)$
		\EndFor
           \State normalize all weights, i.e., set $W_1^n := w_1^n \Big/   \sum_{s=1}^{N_x} w_1^s$  for all $n$
           \For{$t=2, \dots, T$}
		\For{$n=1,\ldots, N_x$}
			\State sample an ancestor $a_{t-1}^n \sim \mathcal{M}(W_{t-1}^1,\dots, W_{t-1}^{N_x})$
			\State draw $x_t^n \sim q_t(\cdot| x_{t-1}^{a_{t-1}^n}, \theta)$
			\State set $w_t^n = p(x_t^n| x_{t-1}^{a_{t-1}^n}, \theta) p(y_t|x_t^n, \theta) \Big/ q_t(x_t^n| x_{t-1}^{a_{t-1}^n}, \theta)$
        	\label{line:unnormalized}
                \EndFor
                \State normalize all weights i.e., set $W_t^n := w_t^n \Big/   \sum_{s=1}^{N_x} w_t^s$    for all $n$
		\State set $\hat p(y_t|y_{1:t-1}, \theta) = (N_x)^{-1} \sum_{n=1}^{N_x} w_t^n$.
	\EndFor
    \Statex
  \end{algorithmic}
\end{algorithm}

Particle filters (PF) are Monte Carlo algorithms that target the filtering distribution $p(x_t|y_{1:t}, \theta)$. Two papers that conducted work in this area are \citet{muller1991monte} and \citet{carpenter1999improved}. More recently, \citet{davey_bayesian_2015} re-create the path of the flight MH370 from satellite data using a state space model estimated with a PF. 
In this Section, we assume that the parameter $\theta$ is fixed and known and we proceed conditionally on this value. The target of PF is the filtering distribution $p(x_T|y_{1:T},\theta)$. As a byproduct, it also computes an approximation of the marginalised likelihood $p(y_{1:T}|\theta)$. The main output is a weighted state sample of size $N_x$ whose empirical distribution
\[\sum_{n=1}^{N_x} W_T^nd_{x_T^n}\]
is an approximation of the filtering distribution. At each stage of the sequential algorithm, PF also produce an estimate of $p(y_t|y_{1:t-1},\theta)$. Then the marginalised likelihood is given by
\[
  p(y_{1:T}|\theta) = p(y_1|\theta) \prod_{t=2}^T p(y_t|y_{1:t-1},\theta).
\]
Importance sampling is at the core of PF:
at time $t=1$ the states are sampled from a proposal distribution $q_1(\cdot|\theta)$, and then the states are sampled from a Markovian kernel $q_t(\cdot | x_{t-1}, \theta)$. The  first reported version of this method \citep{gordon1993novel} sets $q_t$ to the Markovian kernel $p(x_t|x_{t-1}, \theta)$. This proposal may be far from the optimal kernel $p(x_t|x_{t-1},y_t,\theta)\propto p(x_t|x_{t-1},\theta)p(y_t|x_t,\theta)$. More generally, such a particle filter is termed a sequential importance resampling filter by \citet{doucet_sequential_2000}, see Algorithm \ref{algo:PF}. This filter is not necessarily ideal for targeting $p_{t+1}$. Other approaches to particle filtering in literature include a multi-step MCMC kernel \citep{drovandi_estimation_2011}, and the alive particle filter \citep{moral2015alive}.

\subsection{Combined parameter and state inference}\label{sub:SMC2}

We consider now CPS inference, which targets the much more complex distribution $p(x_T, \theta | y_{1:T})$. Such algorithms simultaneously filter the state and compute the posterior of the fixed parameter. We are interested in drawing samples from the joint distribution of $x_T$ and $\theta$ conditional on observations $y_{1:T}$. Initial approaches addressed this problem by appending $\theta$ to the state vector $x_t$, as though the static parameters were dynamic. This approach leads to error that is highly model specific and poorly understood. This led to the development of approaches which explicitly take the fixed nature of $\theta$ into account. \citet{drovandi2016exact} and \cite{drovandi2016alive} address this problem by incorporating the alive particle filter into the PMCMC and SMC algorithms, respectively. \citet{chopin2013smc2} incorporate the sequential importance resampling filter (Algorithm \ref{algo:PF}) into a SMC approach, which they term SMC$^2$ (Algorithm \ref{algo:SMC2}) since particle filters fall within the class of general SMC algorithms. 

\begin{algorithm}
	\caption{SMC$^2$ algorithm \citep{chopin2013smc2} \label{algo:SMC2}}
	\begin{algorithmic}[1]
	\Require $y_{1:T}, N_x, N_{\theta}$
	\Ensure parameter samples $(\theta^1, \dots, \theta^{N_{\theta}})$ with associated normalised weights $(\Omega^1, \dots, \Omega^{N_{\theta}})$
    \For{$m=1,\dots,N_\theta$}
    	\State draw $\theta^m \sim \pi(\theta)$
    	\State start the sequential importance resampling filter given $\theta^m$ to compute $\hat{p}(y_1| \theta^m)$
    	\State set $Z_1^m = \hat{p}(y_1| \theta^m)$
    \EndFor
    \State If $Z_1^m$ are degenerate, rejuvenate $\theta$
    \For{$t=2, \dots, T$}
    	\For{$m = 1, \dots, N_{\theta}$}
    	    \State update the $m$-th sequential filter given $\theta^m$ to compute $\hat{p}(y_t| y_{1:t-1}, \theta^m)$
    	    \State set $Z_t^m = Z_t^{m-1} \hat{p}(y_t| y_{1:t-1}, \theta^m)$
    	\EndFor
    	\State If $\omega$ are degenerate, rejuvenate $\theta$
    \EndFor
	\end{algorithmic}
\end{algorithm}

Algorithm \ref{algo:SMC2} uses the sequential importance resampling filter (Algorithm \ref{algo:PF}) to estimate the likelihood update $\hat{p}(y_t| y_{1:t-1}, \theta^m)$ within a larger SMC algorithm to return weighted parameter samples $(\theta^1, \dots, \theta^{N_{\theta}})$. Since weights $Z_t^m$ are updated rather than used within a resampling step at each time step, rejuvenation is necessary. Otherwise, $Z_t^m$ will become degenerate. So-called sample impoverishment or particle degeneracy is a problem in sequential particle filters. In this paper, we use the classical measure of sample impoverishment called effective sample size \citep[ESS, see, e.g., ][]{djuric2003particle}. This can be estimated by
\begin{align} \label{eq:ESS}
\left( \sum_{m=1}^{N_{\theta}} Z_t^m \right) ^2 \Big/ \sum_{m=1}^{N_{\theta}} (Z_t^m)^2.
\end{align}
Degeneracy is determined based on a fixed ESS threshold. Once degeneracy is detected, the values of $\theta$ are rejuvenated using a weighted kernel density $\sum_{m=1}^{N_{\theta}} Z_t^m K_t(\theta^m, \cdot)$, where $K_t$ is a Markov kernel with invariant distribution $p(\theta|y_{1:t})$ with a particle MCMC \citep{andrieu2010particle}.

\subsection{ABC and state space models}\label{sub:ABCSSM}

The sequential importance resampling particle filter (Algorithm \ref{algo:PF}), which forms part of the SMC$^2$ algorithm (Algorithm \ref{algo:SMC2}), requires computation of the likelihood $p(y_t | x_t^n, \theta)$. This, in many cases, cannot be evaluated but can be used to draw model realisations. In the next section, we discuss approximate Bayesian computation (ABC), a likelihood-free approach to parameter inference that overcomes this limitation. We also review ABC adaptations of PFs and to CPS inference. 


ABC particle filters based on the sequential importance resampling \citep[Algorithm \ref{algo:PF}, see][]{jasra2012filtering,calvet2014accurate,moral2015alive} substitute a likelihood approximator $\hat{p}_{\epsilon}(y_t | x_t^n, \theta)$ for $p(y_t | x_t^n, \theta)$ within Algorithm \ref{algo:PF}. We give now an example of $\hat{p}_{\epsilon}(y_t | x_t^n, \theta)$. Let us assume that we cannot evaluate $p(y_t|x_t, \theta)$, however, knowing $\theta$ and $x_t$, we can simulate a set $\tilde{y}_t^n(i)$, $i=1,\ldots,N_y$, from this distribution. Then we have to replace the evaluation of this density with the proportion of $\tilde{y}_t^n(i)$ that fall near the observation $y_t$ with respect to $d$, i.e,
\begin{align} \label{eq:p.hat.epsilon}
\hat{p}_{\epsilon}(y_t|x_t, \theta) = (N_y)^{-1} \sum_{i=1}^{N_y} \mathbf 1[ d(\tilde{y}_t^n(i), y_t) \le \epsilon]
\end{align}
which is an unbiased estimate of
\begin{align} \label{eq:p.epsilon}
p_{\epsilon}(y_t | x_t^n, \theta) = \int \mathbf 1[ d(\tilde{y}, y_t) \le \epsilon] p(\tilde{y} |x_t, \theta)\text{ d}\tilde{y},
\end{align}
where $\mathbf{1}$ is the indicator function. This could be replaced by any positive-valued decreasing function on $d$, but we use $\mathbf{1}$ for simplicity. Note that $p_{\epsilon}(y_t | x_t^n, \theta)$ as a function of $y_t$ is a proper density function up to a normalising factor that depends only on $\epsilon$. Finally, since we have $N_x$ values of $x_t$ at hand in the PF algorithm at a given time $t$, we have to compute this ABC likelihood estimate $N_x$ times. 

In some cases, the Markovian kernel $p(x_t|x_{t-1},\theta)$ of the state space model is intractable. In these cases, we can set the proposal $q_t(x_t|x_{t-1},\theta^m)$ to $p(x_t|x_{t-1},\theta^m)$ so that they cancel one another in the particle filter weights. This is the approach we take, in order to demonstrate the algorithm for fully intractable problems. In that case, all $u^{m,n}_t$ are equal to $1$, and the ratio of Equation~\eqref{eq:big.ratio} is just the proportion of accepted simulations, weighted by $Z_t^m$, marginalised over parameters and states.

\section{Methodology}  \label{sec:ABCSMC2_methodology}

We discuss now how an ABC particle filter, based on the sequential importance resampling, is incorporated into a likelihood-free CPS algorithm. 

Likelihood-free CPS inference algorithms which have so far been developed \citep{drovandi2016alive,drovandi2016exact}, require that a sequence of thresholds $\epsilon_1, \dots, \epsilon_T$ be chosen before initialisation of the CPS algorithm. However, we develop a technique that bypasses this requirement to allow for automatic $\epsilon$ calibration. 
\subsection{Targets}
We develop this algorithm (Algorithm \ref{algo:ABCSMC2update}) to provide a sample from the joint posteriors $\pi(\theta, x_t | y_{1:t}), t = 1,\dots,T$. However, since some distributions are intractable, we perform a likelihood-free approximation to sample from the approximate joint posteriors defined as follows.
If $t=1$, the target is given by
\begin{gather}
  \label{eq:target_1}
  \pi_{\epsilon_1}(x_1,\theta|y_1)
   =\frac{1}{\pi_{\epsilon_1}(y_1)} 
    \pi(\theta)p(x_1|\theta)
    \int d\tilde{y}_1 p(\tilde y_1|x_1,\theta)\mathbf 1\{d(\tilde y_1,
    y_1)\le \epsilon_1\}, 
  \\
  \text{where} \quad \pi_{\epsilon_1}(y_1)  := 
    \iiint d\theta dx_1 d\tilde{y}_1 \pi(\theta)  p(x_1|\theta)
    p(\tilde y_1|x_1,\theta)\mathbf 1\{d(\tilde y_1,
    y_1)\le \epsilon_1\}.
   \notag
\end{gather}
And, for any $t>1$, the target is defined by induction as
\begin{multline}
  \label{eq:target_t} 
  \pi_{\epsilon_{1:t}}(x_t, \theta|y_{1:t}) :=
  \frac{
  \iint dx_{t-1} d\tilde y_t 
  \pi_{\epsilon_{1:t-1}}(x_{t-1},\theta|y_{1:t-1}) 
  p(x_t|x_{t-1}, \theta)  p(\tilde y_t|x_t,\theta) \mathbf 1\{d(\tilde y_t, y_t)\le \epsilon_t\}}{\pi_{\epsilon_{1:t}}(y_t|y_{1:t-1})},
\end{multline}
where we set
\begin{equation*}
  \pi_{\epsilon_{1:t}}(y_t|y_{1:t-1}) := \iiiint d\theta  \, dx_{t-1} \, 
  dx_t \,  d\tilde y_t \, 
  \pi_{\epsilon_{1:t-1}}(x_{t-1},\theta|y_{1:t-1}) \, 
  p(x_t|x_{t-1}, \theta)  \,   p(\tilde y_t|x_t,\theta)  \, 
  \mathbf 1\{d(\tilde y_t, y_t)\le \epsilon_t\}. 
\end{equation*}
We also introduce, for any $t$,
\[
  \pi_{\epsilon_{1:t}}(y_{1:t}) := \pi_{\epsilon_1}(y_1) \prod_{s=2}^t
  \pi_{\epsilon_{1:s}}(y_s|y_{1:s-1}).
\]


\subsection{Proposed algorithm}
The form of the integral in Equation~\eqref{eq:target_t} suggests to us a strategy for sampling from $\pi_{\epsilon_{1:t}}(x_t, \theta | y_{1:t})$. Let us assume that, at time $t-1$, we have a weighted sample that comprises
\begin{enumerate}
\item a weighted sample of size $N_\theta$, namely $\theta^1, \dots, \theta^{N_\theta}$, with weights $Z_{t-1}^1$, \ldots, $Z_{t-1}^{N_\theta}$, whose weighted empirical distribution is a Monte Carlo approximation of $\pi_{\epsilon_{1:t-1}}(\theta | y_{1:t-1})$; and
\item for each value $\theta^m$ in that sample, a sample $x^{m,1}_{t-1},\dots, x^{m,N_x}_{t-1}$, with weights $W^{m,1}_{t-1}, \dots W^{m,N_x}_{t-1}$, whose weighted empirical distribution is a Monte Carlo approximation of $p_{\epsilon_{1:t-1}}(x_{t-1}|\theta^m, y_{1:t-1})$.
\end{enumerate}
State resampling is performed independently for each value $\theta^m$ of the parameter. Hence, for each $\theta^m$, we still have an approximation of $p_{\epsilon_{1:t-1}}(x_t|\theta, y_{1:t-1})$. For each value $\theta^m$, we resample the states using their weights $W^{m,n}_{t-1}$ with the multinomial distribution $\mathcal{M}$. Note that other sampling choices, such as systematic sampling \citep{li2015resampling}, can be contemplated and incorporated in a straightforward manner. We move the resampled states according to $q_t$ to get the $x^{m,n}_t$ proposals. And we draw $N_y$ simulations $\tilde{y}_t^{m,n}(i)$, $i=1,\ldots,N_y$, for each $x^{m,n}_t$, according to the emission distribution $p(\tilde{y} | x^{m,n}_t, \theta^m)$.
Finally, the unnormalised weight $w^{m,n}_{t}$ of $x^{m,n}_t$ is
\[
  w^{m,n}_t := u^{m,n}_t \sum_{i=1}^{N_y} \mathbf 1 \left\{ d[\tilde{y}_t^{m,n}(i),y_t] \le \epsilon_t \right\}, \quad \text{where }u^{m,n}_t := p\left( x^{m,n}_t\Big|x^{m,a}_{t-1},\theta^m \right) \Big/
  q_t\Big( x^{m,n}_t\Big|x^{m,a}_{t-1},\theta^m \Big),
\]
where $a$ denotes the index of the ancestor from the multinomial resampling. This allows us to estimate $\hat{p}_{\epsilon_{1:t}}(y_t| y_{1:t-1}, \theta^m)$ by taking an average $(N_x)^{-1} \sum_n w^{m,n}_{t}$ from the particle approximation, where $\hat{p}_{\epsilon_{1}}(y_1 | y_{1:0}, \theta^m) := \hat{p}_{\epsilon_{1}}(y_1 | \theta^m)$. 

We discuss now the calibration of $\epsilon$. In the standard application of ABC rejection, it is usual to calibrate $\epsilon$ as a quantile ``acceptance probability" $P_{\text{acc}}$ of the distances of the simulated values $\tilde{y}$, see Section~\ref{sub:ABC}. To achieve this, the particles $x^{m,n}_t$ can be ordered according to their distances $ d^{m,n}_t(i) = d(\tilde{y}_t^{m,n}(i),y_t)$, $i=1,\dots, N_y$, and the smallest value of $\epsilon_t$ can be found such that
\begin{equation}
   \label{eq:big.ratio} 
  \frac{\displaystyle\sum_m^{N_{\theta}} Z_t^m \sum_n^{N_x} u^{m,n}_t \sum_{i=1}^{N_y} \mathbf 1 \left\{ d^{m,n}_t(i) \le \epsilon_t \right\}}{\displaystyle
    \sum_m^{N_{\theta}} Z_t^m \sum_n^{N_x} u^{m,n}_t N_y}
\end{equation}
is greater than $P_{\text{acc}}$. 

\begin{algorithm}
\caption{Self-calibrated ABCSMC$^2$ update \label{algo:ABCSMC2update}}
\begin{algorithmic}[1]

\Require $\left( \theta^m, \omega^m, Z^m_{t-1}, x^{m,1:N_x}_{t-1}, W^{m,1:N_x}_{t-1} \right)$, $m=1,\ldots, N_\theta$; new observation $y_t$; the ABC acceptance $P_{\text{acc}}$; and optionally, previously computed thresholds $\epsilon_{1:{t-1}}$.
\Ensure $\left( \theta^m, \omega^m, Z^m_t, x^{m,1:N_x}_{t}, W^{m,1:N_x}_{t} \right)$, $m=1,\ldots, N_\theta$; and computed thresholds $\epsilon_{1:t}$.
\Statex

\For{$m=1,\ldots,N_\theta$}
	\For{$n=1,\ldots, N_x$}
		\State draw an ancestor $a^{m,n}_{t-1} \sim \mathcal{M}\Big(W^{m,1}_{t-1},\dots, W^{m,N_x}_{t-1}\Big)$
		and $x^{m,n}_{t} \sim q_t(\cdot| x^{m,a^{m,n}_{t-1}}_{t-1}, \theta^m)$
		\For{$i=1, \dots, N_y$}
			\State draw $\tilde{y}_t^{m,n}(i) \sim p(\tilde{y} | x^{m,n}_{t}, \theta^m)$
			and set $d^{m,n}_t(i) = d(\tilde{y}_t^{m,n}(i),y_t)$ 
		\EndFor
		\State set $u^{m,n}_t = p\left( x^{m,n}_t\Big|x^{m,a^{m,n}_{t-1}}_{t-1},\theta^m \right) \Big/
  q_t\Big( x^{m,n}_t\Big|x^{m,a^{m,n}_{t-1}}_{t-1},\theta^m \Big)$
	\EndFor
\EndFor
\If{$\epsilon_t$ was not included as input}
\State find the smallest $\epsilon_t$ such that Equation~\eqref{eq:big.ratio} is greater than $P_{\text{acc}}$
\EndIf

\For{$m=1,\ldots, N_\theta$}
	\For{$n=1,\ldots, N_x$}
		\State set $w^{m,n}_{t} = u^{m,n}_t (N_y)^{-1}\sum_{i=1}^{N_y} \mathbf 1 \left\{ d[\tilde{y}_t^{m,n}(i),y_t] \le \epsilon_t \right\} $ 
	\EndFor
	\State set $\hat{p}_{\epsilon_t}(y_t | y_{1:t-1}, \theta^m) = (N_x)^{-1} \sum_{n=1}^{N_x} w_t^{m,n}$
	\If{$\hat{p}_{\epsilon_t}(y_t | y_{1:t-1}, \theta^m)>0$}
		\State normalise the weights by setting $W^{m,n}_t = w^{m,n}_{t}\big/\big(\sum_{n'} w_t^{m,n'}\big)$ for all $n=1,\ldots, N_x$
	\Else
		\State set $W^{n,m}_t=0$ for all $n=1,\ldots, N_x$
	\EndIf
	\State set $Z_t^m = Z_{t-1}^m \times \hat{p}_{\epsilon_t}(y_t | y_{1:t-1}, \theta^m)$
\EndFor
\If{degeneracy}
	\For{$m=1,\ldots, N_\theta$}
		\State draw $\alpha^m \sim \mathcal{M}\Big(Z^{1}_{t},\dots, Z^{N_{\theta}}_{t}\Big)$
		\State reindex $ \left( \theta^m, Z^m_t, x^{m,1:N_x}_{t}, W^{m,1:N_x}_{t} \right) = \left( \theta^{\alpha^m}, Z^{\alpha^m}_t, x^{{\alpha^m},1:N_x}_{t}, W^{{\alpha^m},1:N_x}_{t} \right)$
		\State draw $\breve{\theta}^m \sim K_t(\cdot|\theta^m)$
		\State run a new set of self-calibrated ABCSMC$^2$ updates (This algorithm, Algorithm \ref{algo:ABCSMC2update}),
                \State \quad $|$\ until time $t$, with for $\breve{\theta}^m$ previously calibrated $\epsilon_{1:t}$ and observations $y_{1:t}$.
		\State reindex $\left( \theta^m, Z^m_t, x^{m,1:N_x}_{t}, W^{m,1:N_x}_{t} \right) = \left( \breve{\theta}^m, \breve{Z}^m_t, \breve{x}^{m,1:N_x}_{t}, \breve{W}^{m,1:N_x}_{t} \right)$ with probability
\[
1 \wedge \frac{\pi(\breve\theta)\, \breve{Z}_t K_t(\breve{\theta}^m|\theta^m)}{\pi(\theta) Z_t K_t(\theta^m|\breve{\theta}^m)}.
\]
		\State set $Z_t^m = 1$
	\EndFor
\EndIf		
\end{algorithmic}
\end{algorithm}


With each ABCSMC$^2$ update, more data $y_t$ are incorporated into the static parameter posterior $\pi_{\epsilon_{1:t}}(\theta|y_{1:t})$. This leads to degeneracy of the weights $Z_t^m$. To check degeneracy, we use the ESS criteria in Equation~\eqref{eq:ESS}, with threshold chosen prior to analysis. In order to propose new parameters and further improve the particle approximation to $\pi_{\epsilon_{1:t}}(\theta|y_{1:t})$, we need an estimate of the ABC marginal likelihood $ p_{\epsilon_{1:t}}(y_{1:t}|\theta^m)$. Following \citet{chopin2013smc2}, the ABC likelihood can be estimated, in the following manner:
\begin{align*}
\hat{p}_{\epsilon_{1:t}}(y_{1:t}|\theta^m) :=\prod_{s=1}^t \hat{p}_{\epsilon_{1:s}}(y_s| y_{1:s-1}, \theta^m).
\end{align*}
Once we have $Z^m_t$ for each $\theta^m$ we can rejuvenate the particle system. The steps involved in parameter rejuvenation are the following:
\begin{enumerate}
\item resample $\theta^m$, $m=1,\dots, N_{\theta}$ to make $Z_t^m$ non-degenerate,
\item generate new parameter proposals $\breve{\theta}^m$ for each $m$, from a kernel $K_t$ around $\theta^m$,
\item for each $m$ run an independent ABC particle filter, conditional on $\breve{\theta}^m$, with the previously computed thresholds $\epsilon_{1:t}$, in order to compute the particle system at time $t$, and 
\item accept each $\left( \breve{\theta}^m, \breve{Z}^m_t, \breve{x}^{m,1:N_x}_{t}, \breve{W}^{m,1:N_x}_{t} \right)$, $m = 1, \dots, N_{\theta}$ with probability
\[
1 \wedge \frac{\pi(\breve\theta)\, \breve{Z}_t K_t(\breve{\theta}^m|\theta^m)}{\pi(\theta) Z_t K_t(\theta^m|\breve{\theta}^m)}.
\]  
\end{enumerate}
It is important to recycle the previously computed thresholds $\epsilon_{1:t}$ so that the rejuvenation step preserves the distribution of the current particle set, i.e.\ from the approximate posterior $\pi_{\epsilon_{1:t}}(\theta^m|y_{1:t})$. 

The rejuvenation step means that the entire state space model, up to time $t$, is re-evaluated, for proposed parameters $\breve{\theta}$, within the ABC particle filter whenever degeneracy occurs. This means that the computational complexity of our algorithm is not linear with time. This is also true for the algorithm of \citet{chopin2013smc2}. 


\subsection{Theoretical results}
In this Section, we set the following hypotheses.
\begin{enumerate}
\item[\bf (A1)] The sampling distributions are chosen such that, for all
  $\theta$ and $x_{t-1}$,
  \[
    p(\cdot|x_{t-1},y_t,\theta) \ll q_t(\cdot|x_{t-1},\theta).
  \]
\item[\bf (A2)] The $\theta^m$, $m=1,\ldots, N_\theta$, are never resampled. 
\item[\bf  (A3)] All $\epsilon_t$ have been set a priori.
\end{enumerate}
The first assumption \textbf{(A1)} is quite classical and is necessary so that the importance sampling at each stage is unbiased. 
The second assumption \textbf{(A2)} is set to simplify the results below since the effects of the $\theta^m$'s resampling are well known. The last assumption will be alleviated in the last Proposition of this Section to understand the consequences of the threshold calibration.

The first results, in both following Propositions, show that the un-normalised ABC-SMC$^2$ estimate is unbiased and consistent. Their proofs are given in Appendices~\ref{sub:proof_expected_value} and \ref{sub:proof_LLN}, respectively. 
\begin{pro} \label{pro:expected_value}
  Assume \textbf{(A1)}, \textbf{(A2)} and \textbf{(A3)} hold. Consider a bounded function $\varphi(x_t, \theta)$ and set
  \[
    \check\varphi_t = \frac{1}{N_\theta}\sum_{m=1}^{N_\theta}
    Z_t^m \sum_{n=1}^{N_x}
    W_t^{m,n} \varphi(x_t^{m,n},\theta^m). 
  \]
  We have
  \begin{equation} \label{eq:expected_value}
    \esp(\check\varphi_t) = \pi_{\epsilon_{1:t}}(y_{1:t}) \iint d\theta
    \, dx_t \,
    \pi_{\epsilon_{1:t}}(x_t, \theta|y_{1:t})  \, \varphi(x_t,\theta).
  \end{equation}
\end{pro}
\begin{pro} \label{pro:LLN}
  With the same assumptions and notations of Proposition 1. When $N_\theta\to \infty$,
  \[
    \check\varphi_t  \to \pi_{\epsilon_{1:t}}(y_{1:t}) \iint d\theta
    \, dx_t \,
    \pi_{\epsilon_{1:t}}(x_t, \theta|y_{1:t})  \, \varphi(x_t,\theta) \quad\text{a.s.}
  \]
\end{pro}

\bigskip

The ABC-SMC$^2$ estimate of the expected value of a bounded function $\varphi(x_t,\theta)$ with respect to the target \eqref{eq:target_t} is \[
  \hat \varphi_t = \check \varphi_t \Big / \Omega_t,
  \quad \text{where }
  \Omega_t := \frac{1}{N_\theta}\sum_{m=1}^{N_\theta}
  Z_t^m \sum_{n=1}^{N_x}
  W_t^{m,n},
\]
and $\check \varphi_t$ is defined as in
Propositon~\ref{pro:expected_value}.
We can now state the main result:
\begin{theo} \label{theo:LLN}
 Assume \textbf{(A1)}, \textbf{(A2)} and \textbf{(A3)} hold. When $N_\theta\to\infty$,
  \[
    \hat\varphi_t \to \iint d\theta
    \, dx_t \,
    \pi_{\epsilon_{1:t}}(x_t, \theta|y_{1:t})  \, \varphi(x_t,\theta)
    \quad \text{a.s.}
  \]
\end{theo}
This Theorem is a direct consequence of Proposition~\ref{pro:LLN} since $\Omega_t$ is a special case of $\check\varphi_t$ when the bounded function $\varphi(x_t,\theta) \equiv 1$.

\bigskip

The theoretical results above show that the scheme is consistent even if $N_x=N_y=1$. But the numerical accuracy at finite sample sizes depends on the values of $N_x$ and $N_y$. In particular, at each stage, the total number of $\tilde y_t$'s simulations is the product $N_x N_y$. A larger number of these $y_t|\theta^m$ samples improves the calibration for the sequence of thresholds $\epsilon_{1:T}$. See, for example, \citet{del2012adaptive} which also discussed a similar issue in Section~4.1.2 of their paper.

Assume now that \textbf{(A1)} and \textbf{(A2)} hold and that $\epsilon_1$,\ldots,$\epsilon_{t-1}$ have been set a priori. 
With Propositions~\ref{pro:expected_value} and \ref{pro:LLN}, for any value of $\epsilon_t>0$, the numerator of Equation~\eqref{eq:big.ratio} is an unbiased, consistent estimate of
\begin{align*}
  p_{\epsilon_{1:t}}(y_{1:t})
  &= p\Big( A_{\epsilon_{1:t}}\Big)
  \\
  & = p_{\epsilon_{1:t-1}}(y_{1:t-1})\, p\Big( d(\tilde y_t, y_t)\le\epsilon_t \Big| A_{\epsilon_{1:t-1}} \Big), \quad \text{where}
  \\
  & \quad A_{\epsilon_{1:s}} = \bigcap_{r=1}^s \Big\{ d(\tilde y_r, y_r)\le \epsilon_r \Big\} \quad \text{for any }s.
\end{align*}
Likewise, the denominator of Equation~\eqref{eq:big.ratio} is an unbiased, consistent estimate of the same probability, but with $\epsilon_t=\infty$. Thus, it estimates fairly $p_{\epsilon_{1:t-1}}(y_{1:t-1})$, and the ratio in Equation~\eqref{eq:big.ratio} is a consistent estimate of $p\Big( d(\tilde y_t, y_t)\le\epsilon_t \Big| A_{\epsilon_{1:t-1}} \Big)$. Hence, the rule we propose to tune $\epsilon_t$ is to choose a Monte-Carlo estimate of the quantile of order $P_\text{acc}$ of the distance of $\tilde y_t$ to $y_t$ given the information we retrieved until time $t-1$. Note that this probability is averaged over all values of $\theta$ from the ABC posterior recovered until time $t-1$.

The estimate of the update of the marginal likelihood in Algorithm~\ref{algo:ABCSMC2update}, namely $\hat p_{\epsilon_t}(y_t|y_{1:t-1},\theta^m)$, is the relative weights of simulations drawn from $\theta^m$ that satisfy the inequality $d(\tilde y_t^{m,n}(i), y_t)\le \epsilon_t$. So, even if $\epsilon_t$ is calibrated on the whole set of simulations at stage $t$, this estimate is a fair ABC surrogate of the update of the likelihood. In the extreme case of $N_x=N_y=1$ and a proposal $q_t(x_t|x_{t-1},\theta)=p(x_t|x_{t-1},\theta)$, this estimate is
\begin{align*}
  \hat p_{\epsilon_t}(y_t|y_{1:t-1},\theta^m)
  &=
    \frac{\ds \mathbf 1\{ d(\tilde y_t^{m,1}(1), y_t)\le \epsilon_t\}}{\ds
    \sum_{m'=1}^{N_\theta} \mathbf 1\{ d(\tilde y_t^{m',1}(1), y_t)\le \epsilon_t\} }
  \\
  & =
    \begin{cases}
      1\big/(N_\theta P_\text{acc}) & \text{if }\tilde y_t^{m,1}(1)\text{ has been accepted},\\
      0 & \text{otherwise}.
    \end{cases}
\end{align*}
This is the standard ABC approximation of the likelihood. Increasing $N_x$ or $N_y$ sharpens the approximation of the likelihood.

\section{Examples} \label{sec:ABCSMC2_examples}

We now demonstrate the use of our self-calibrated ABCSMC$^2$ update (Algorithm~\ref{algo:ABCSMC2update}) with some examples.

\subsection{Skew normal distribution}
\label{sub:skew}

\begin{figure}
    \centering
    \includegraphics[width=.6\textwidth]{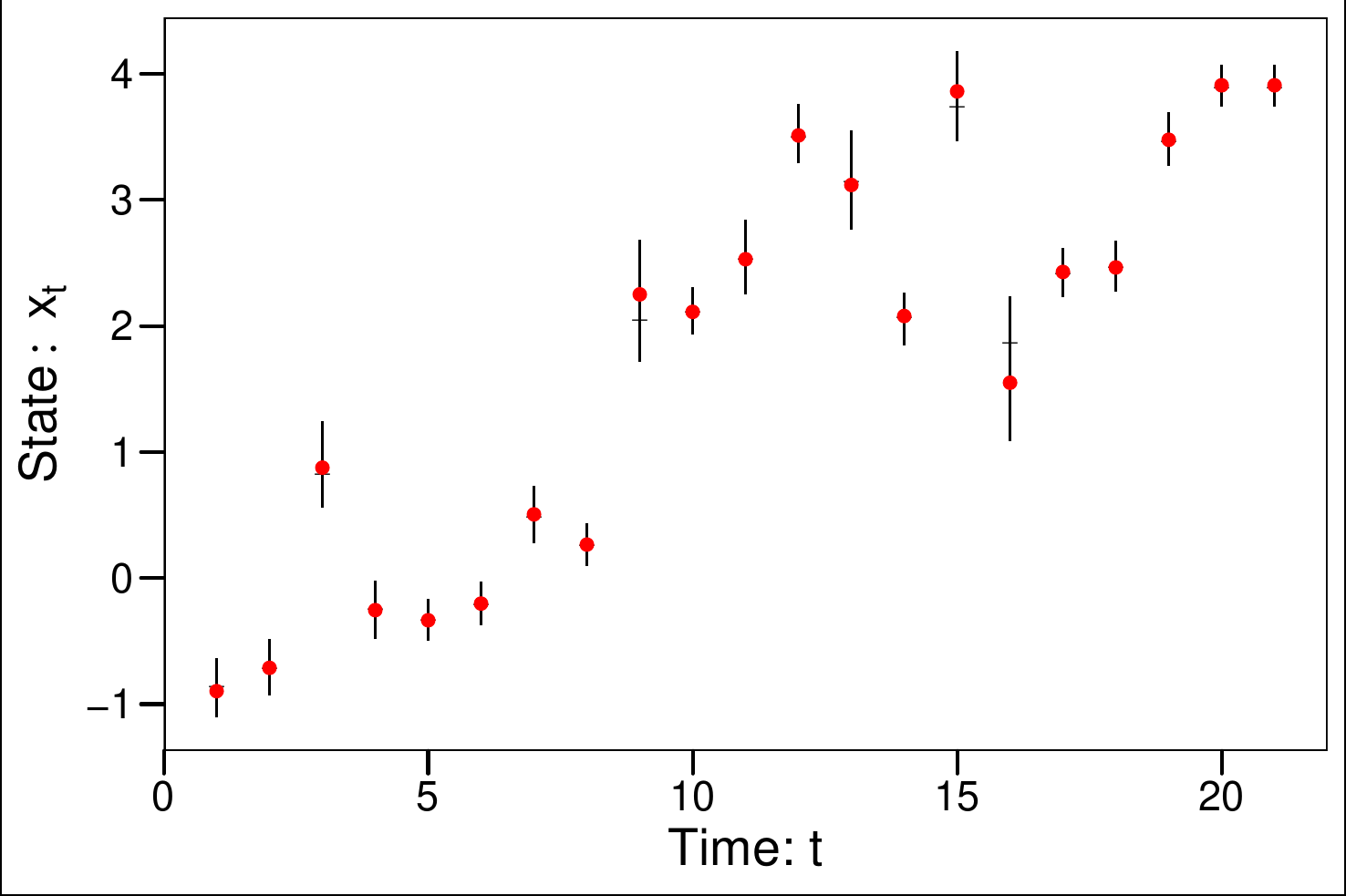}
    \caption{Skew normal example. The true latent state $x_t$ is in red and the Bayesian 95\% prediction intervals of $p(x_t|y_{1:t})$ are the black vertical lines. The black horizontal line is the empirical median of $p(x_t|y_{1:t})$.}
    \label{fig:skew_state}
\end{figure}

\begin{figure}
    \centering
    \includegraphics[width=.9\textwidth]{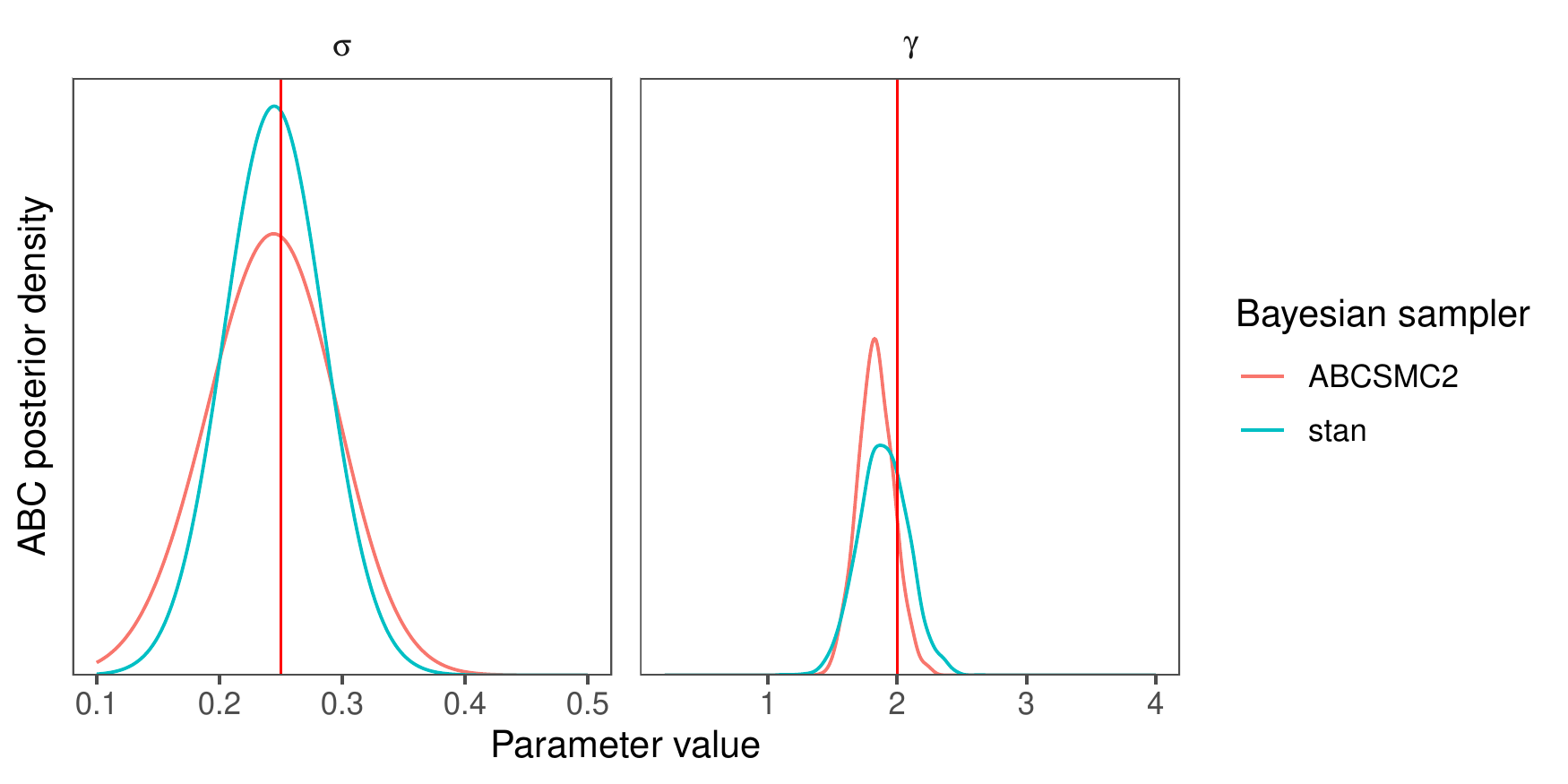}
    \caption{Skew normal example. Marginal approximation of posterior densities $p(\sigma | y_{1:20})$ and $p(\gamma | y_{1:20})$: from our ABCSMC$^2$ in red and; from STAN sampler in blue. The true values are at the vertical solid red lines. The priors used for parameter $\sigma$ and $\gamma$ are $U(0.1, 0.5)$ and $U(0.2, 4)$ respectively.}
    \label{fig:skew_parameter}
\end{figure}

\subsubsection{Method}
As a demonstration of our approach, we consider a simple state space model driven by an auto-regressive process. The emission distribution is the skewed version of the normal distribution as defined by \citet{azzalini1985class}. The skew normal distribution can be parameterised in a similar way to the normal distribution, with an additional parameter $\gamma$ for skewness (see \citet{azzalini2005skew} for further details); we represent this distribution as SN($\mu, \sigma, \gamma$). The statistical model is as follows: $\theta=(\sigma, \gamma)$,
\begin{align*}
x_{t} | x_{t-1}, \theta&\sim \mathrm{N}(x_{t-1},1) \\
(y_{t,1} ~, \dots, ~ y_{t,10}) | x_{t}, \theta &\overset{\text{iid}}{\sim} \mathrm{SN}(x_{t}, \sigma, \gamma),
\end{align*}
with initial state distribution $p(x_1) = \mathrm{N}(0, 1)$, and where each observation $y_t$ comprises 100 independent and identically distributed observations of the skew normal distribution at each time step $t$. The state parameter $x_{t}$ is the mean parameter of the SN distribution, and the parameters to be inferred are $\sigma$ and $\gamma$. The priors are $\sigma \sim U(0.1, 0.5)$ and $\gamma \sim U(0.2, 4)$. The true parameters used for the simulations are $\sigma = 0.25$ and $\gamma = 2$. 

The summary statistic $s_t:=\eta(y_t)$ for this example is composed of the sample mean $\bar y_t$, standard deviation $\hat\sigma(y_t)$ and skewness $B(y_t)$. Skewness is calculated using a method discussed in \citet{joanes1998comparing}: $ B(y_t) = m_3(y_t)\big/ \hat\sigma^3(y_t)$,
as implemented by the R package \texttt{e1071} \citep{e1071}, where $m_3(y_t)$ is the sample third central moment. The distance is the weighted Euclidean distance between observed and simulated summary statistics. We implement this model with $N_{\theta} = 2,000$, $N_{x} = 2,000$, $N_y = 100$, and $P_{\text{acc}} = 0.05$ for time $t = 1, \dots, 20$. The ESS threshold to rejuvenate the $\theta$-particles is set to $0.5 N_{\theta}$. 

This example was intentionally made short on the time scale to make validation, with a Hamiltonian Montel Carlo likelihood-based sampler written in STAN \citep{carpenter2017stan}, possible. The point of this is to validate our approach rather than to compare performance. 


\subsubsection{Results}
Inference results for the marginalised filtering distribution $p(x_t|y_{1:t})$ for $t = 1, \dots, 20$, as shown in Figure~\ref{fig:skew_state}. We can see that the algorithm performs well in matching the true state. The posterior $p(\theta | y_{1:t})$ for the static parameters is shown in Figure \ref{fig:skew_parameter}. The results here are also good. The $\sigma$ parameter is estimated well, with posterior mass concentrated near the true parameter, similarly for the $\gamma$ parameter displays more posterior variance. We included a comparison with the likelihood-based Hamiltonian Monte Carlo Bayesian probabilistic programming language, stan. Of course the likelihood-based method outperforms our method, but the fact that our approximate posterior is close to the true posterior is highly encouraging. 



\subsection{Econometric model}
\label{sub:svm}

\begin{figure} 
    \centering
    \includegraphics{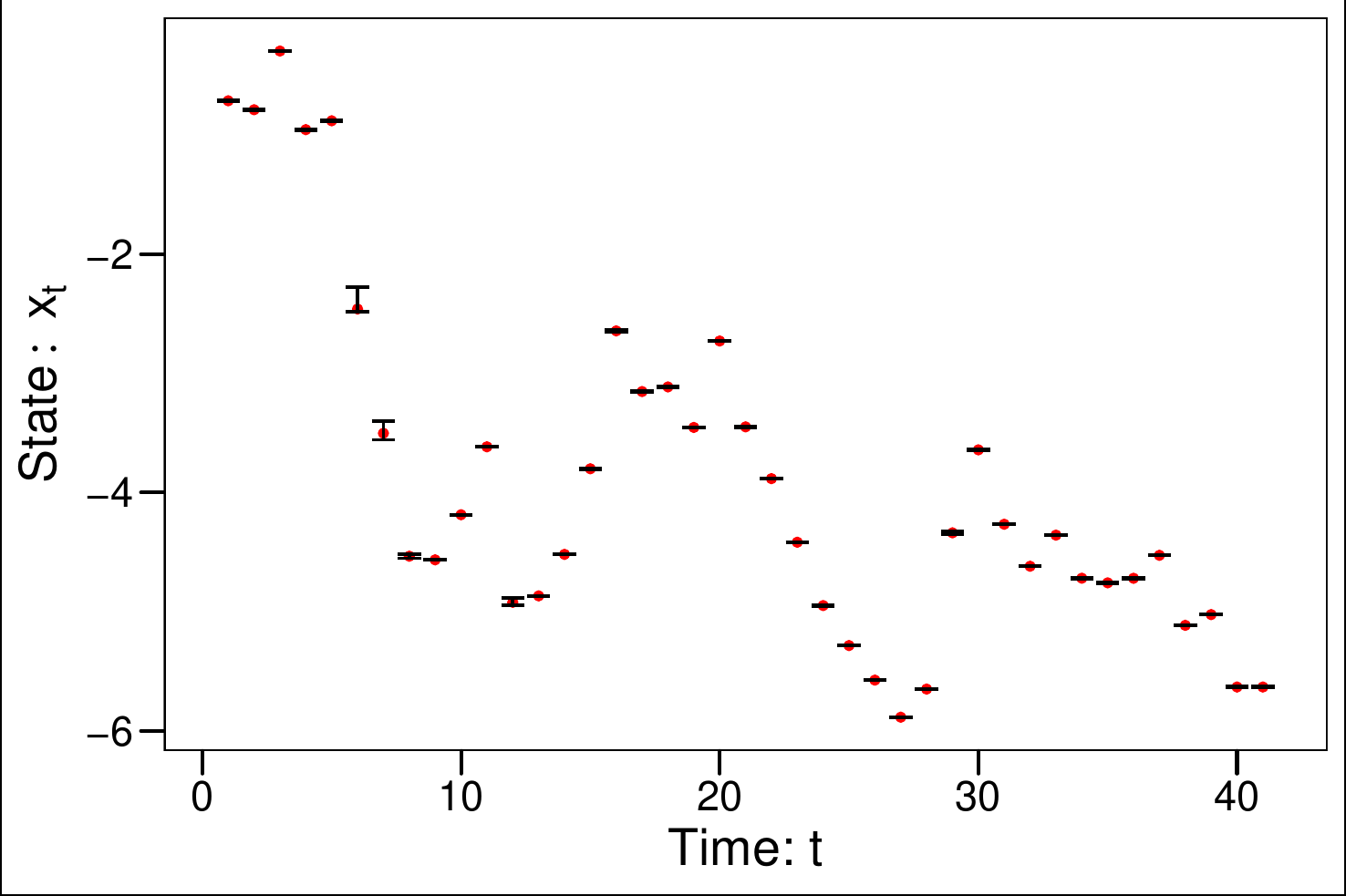}
    \caption{Econometric example. True latent state $x_t$ in red with Bayesian 95\% prediction intervals of $p[x_t|y_{1:40}]$ as error bars. }
    \label{fig:Vol_state}
\end{figure}

\begin{figure}
    \centering
    \includegraphics{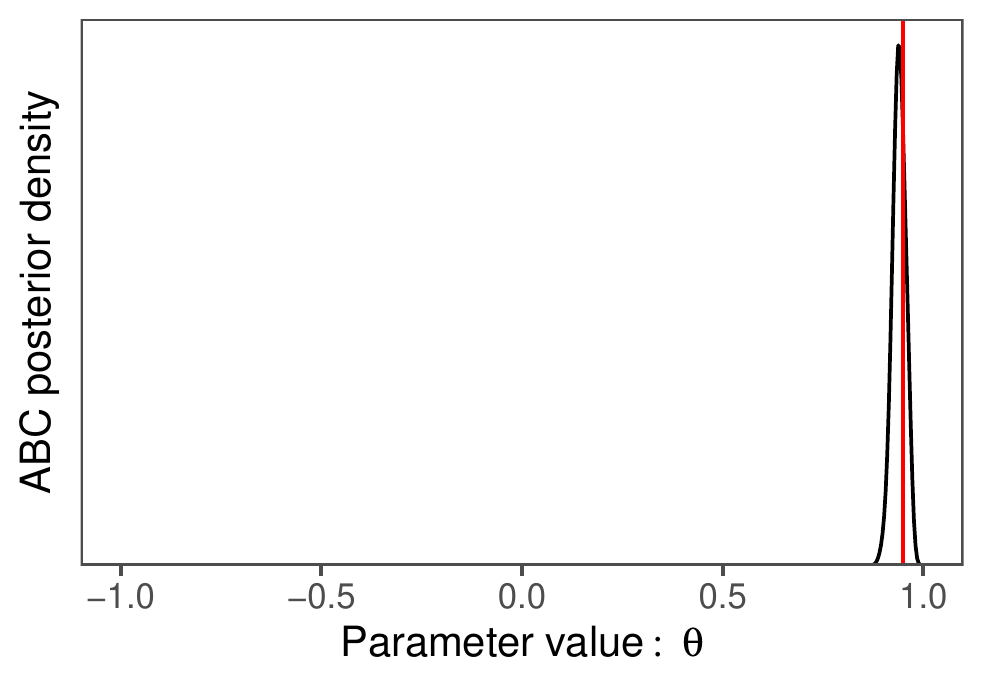}
    \caption{Econometric example. Marginal ABCSMC$^2$ posterior density $p(\theta | y_{1:40})$, in black, compared with true value (vertical solid red line). The prior for the parameter is $U(-1, 1)$.}
    \label{fig:Vol_parameter}
\end{figure}

\begin{figure}
    \centering
    \includegraphics{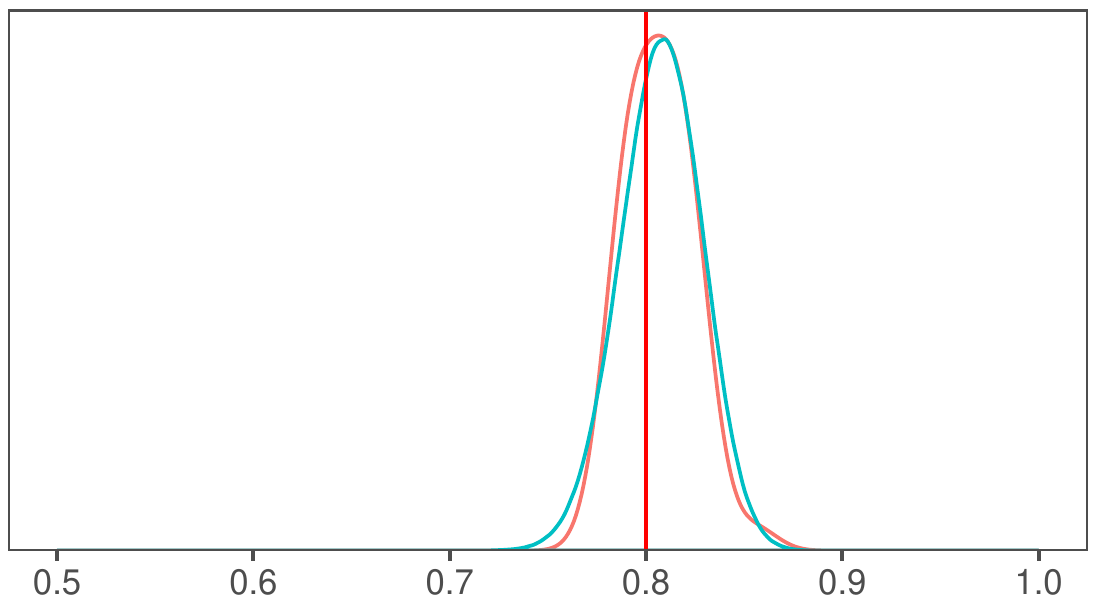}
    \caption{Econometric example. ABCSMC$^2$ posterior density $p(\theta | y_{1:20})$ in red compared to the output of the Stan sampler in blue. The prior for the parameter is $U(-1, 1)$.}
    \label{fig:Vol_parameter_lik}
  \end{figure}

{Stochastic volatility models}  are one of the most common state space models in the econometric literature. Recently, \citet{vankov2019filtering} considered the case where the emission distribution is a stable distribution, that can take into account heavily tailed volatility. Assuming known values for static parameters of the model, they propose a likelihood-free particle filter to infer the states. But, up to our knowledge, no specific algorithm is available to take into account the uncertainty on the static parameters of the model. The model is as follows: 
\begin{align*}
    x_{t}|x_{t-1},\theta &\sim \mathrm{N}(\mu + \theta x_{t-1}, \sigma_h) \\
    v_{t} & \sim \mathrm{SD}(\alpha, \beta, \gamma, \delta) \\
    y_{t} &= \exp(0.5 x_{t}) v_{t}
\end{align*}
where $\mathrm{SD}(\alpha, \beta, \gamma,\delta)$ is the stable distribution \citep{lombardi2009indirect, mandelbrot1963variation}, with characteristic function
\[
F(u|\alpha,\beta,\gamma,\delta) = \begin{cases} \displaystyle
\exp\left\{  - \gamma^\alpha |u|^\alpha \left[1-i \beta \tan\frac{\pi\alpha}{2}(\text{sign}\,u) +i\delta u\right]
\right\} & \text{if }\alpha\neq 1,
\\
\\
\displaystyle
\exp\left\{ -\gamma |u|\left[1+i\beta \frac{2}{\pi} (\text{sign}\, u)\log|u| +i\delta u\right]
\right\}
& \text{if } \alpha=1.
\end{cases}
\]
Note that $\alpha$ is a stability parameter, $\beta$ a skewness parameter, and $\gamma$ and $\delta$ are related to the scale and the position of the distribution. In particular, if $\alpha=2$ and $\beta=0$, the stable distribution is a Gaussian distribution; if $\alpha=1$ and $\beta=0$, it is a Cauchy distribution; if $\alpha=0.5$ and $\beta=1$, it is a Lévy distribution. Apart from these examples, the density of stable distribution is intractable \citep{lombardi2007bayesian}. This has led to the development of ABC techniques for stable distributions \citep{peters2012likelihood}. 

\subsubsection{Method}
All static parameters are known except for $\theta$. The prior of $\theta$ is $U(-1, 1)$. We define a transformation on the variables $\xi := \Phi^{-1} \left(\frac{\theta + 1}{2} \right)$, where $\Phi^{-1}$ is the inverse CDF of the standard normal distribution. The transition kernel $K_t$ (Algorithm \ref{algo:ABCSMC2update}) is the following:
\begin{align}
    \tilde{\xi} &\sim N\left[\xi, c\hat{\sigma}_{\xi} \right],
\end{align}
where $\hat{\sigma}_{\xi}$ is the weighted standard deviation of $\xi$ and $c$ is a fixed tuning parameter set to 0.1 to ensure high acceptance. The ratio which appears in the acceptance probability of Algorithm \ref{algo:ABCSMC2update} is:
\begin{align}
\frac{K_t(\breve{\theta}^m|\theta^m)}{K_t(\theta^m|\breve{\theta}^m)} &= \frac{\phi\left[\Phi^{-1}(\frac{\theta^m + 1}{2})\right]}{\phi\left[\Phi^{-1}(\frac{\breve{\theta}^m + 1}{2})\right]},
\end{align}
where $\phi$ is the standard normal probability density function, because the random walk is on the tranformed variable $\xi$.

\subsubsection{Results}
The tuning parameters were $N_{\theta} = 80,$ $N_x = 5 \times 10^4$, $N_y = 1$. The acceptance probability to automatically tune the threshold was $P_{\text{acc}} = 0.005$. The ESS threshold to rejuvenate the $\theta$-particles was $0.5 \, N_{\theta}$. The marginal filtering distribution follows the true state closely (Figure~\ref{fig:Vol_state}), despite taking into account the uncertainty on $\theta$. The same is true for the posterior density in Figure~\ref{fig:Vol_parameter}.

Very few alternatives exist to estimate the parameters of the stochastic volatility model including a general stable distribution because its density is intractable. But we can assume that we are in the special case where this distribution is Gaussian, so that the completed likelihood can be evaluated. We adapt the STAN sampling algorithm of \citet{stan_stochvol} to sample from the true posterior on a smaller dataset of size $T=20$.
Figure~\ref{fig:Vol_parameter_lik} compares the posterior distribution on $\theta$ recovered by STAN and the one we get with our method. We clearly see that the approximate posterior takes the form of the true posterior, which provides an additional check as to the validity of our approach.

In this example, each $y_t$ is of size $1$, and we did not need to summarise it to run our algorithm. The whole dataset $y=(y_1,\ldots, y_T)$ is of size $T=40$ (general case) or of size $T=20$ (Gaussian case). Due to the curse of dimensionality, it would not have been possible to get decent results with a basic accept-reject ABC algorithm without introducing a summary statistic of much smaller dimension, thus without losing part of the information given by the data. Figure~\ref{fig:Vol_parameter_lik} is a demonstration that likelihood-free methods can return sharp results if we are able to run the algorithm without the resort to a non linear projection of the datasets, and without facing the curse of dimensionality. 


\subsection{Semi-parametric Hawkes process}
\label{sub:hawkes}

Hawkes processes, as introduced by \citet{hawkes1971point,hawkes1971spectra} are a type of self-exciting point process on $\mathbb R_+$. The term \textit{self-exciting} means that events are not independent. Assume that $\upsilon_i$, $i\in\mathbb N$ are the ordered events. All previous events $\upsilon_i$, $i\le j$ raise the probability a new event $\upsilon_{j+1}$ occurring. The rate for a Hawkes process at time $\tau$, given all past events $\{\upsilon_j:\ \upsilon_j\leq \tau\}$, is
\[
  \lambda(\tau) = \lambda_b(\tau) + \sum_{\upsilon_j \leq \tau} K_H(\upsilon_j,\tau).
\]
We use $\tau$ rather than $t$ to emphasise that this time variable is continuous. Here $\lambda_b$ is some baseline rate function that is conditionally independent of $\upsilon_j$; and $K_H$ is the self-exciting kernel function evaluated over $\upsilon_j$ and time $\tau$. We describe now the specifics of the Hawkes process we intend to study. 


\subsubsection{Method}
%

\begin{figure}
    \centering
    \includegraphics[width=.72\textwidth]{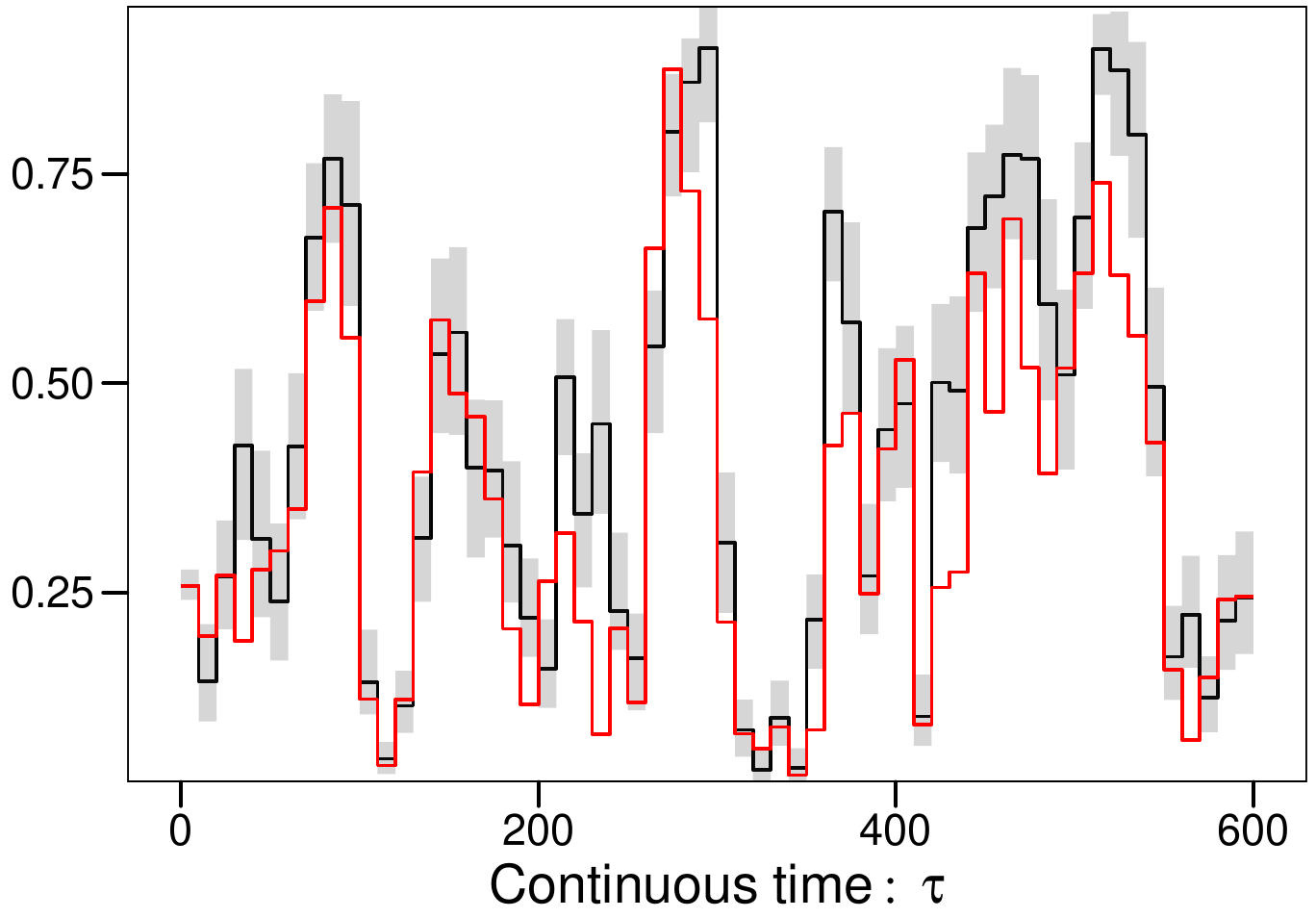}
    \caption{Hawkes example. True transformed latent state $\ell_t=\text{logit}^{-1}(L_t)$ in red with Bayesian 95\% prediction intervals of $p[\text{logit}^{-1}(L_t)|y_{1:t}]$ in grey. The solid black line is the empirical median of $p[\text{logit}^{-1}(L_t)|y_{1:t}]$.}
    \label{fig:Hawkes_state}
\end{figure}

\begin{figure}
    \centering
    \includegraphics{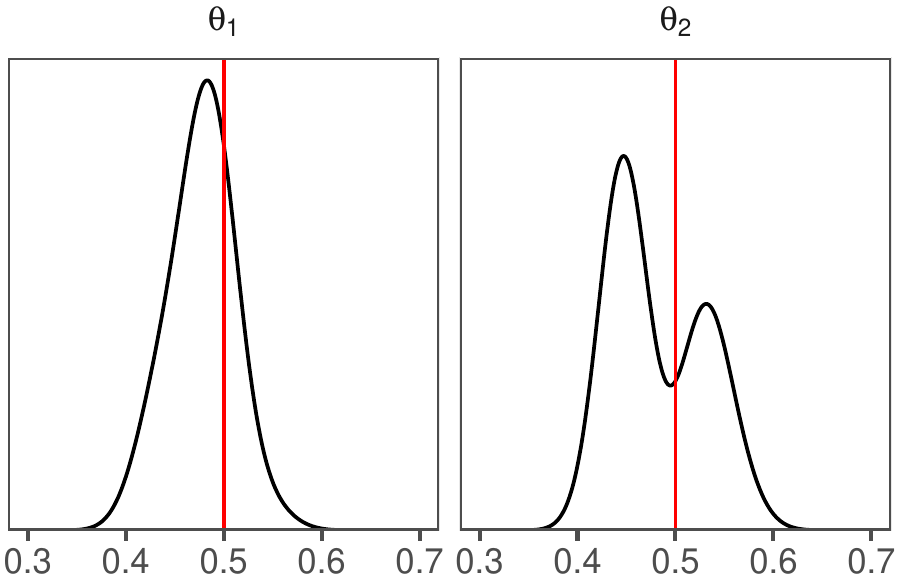}
    \caption{Hawkes example. Marginal ABC posterior densities $p(\theta_1 | y_{1:60})$ and $p(\theta_2 | y_{1:60})$, in black, compared with true values (vertical solid red lines). The prior use for each parameter was $U(0.3, 0.7)$.}
    \label{fig:Hawkes_parameter}
\end{figure}

We set a simple, nonparametric prior on the baseline rate $\lambda_b$. To that aim, we assume that this intensity is a step function.   The time axis is divided into non-overlapping intervals, $I_t=[10(t-1); 10t)$, $t=1,2,\ldots$ and the levels of the baseline intensity are driven by an auto-regressive process of order 1, AR(1) and a coefficient $\alpha$. In other words,
\begin{align*}
  \lambda_b(\tau) &= \theta_0 \sum_{i=1}^{n_i} \ell_t~ \mathbf 1\{\tau \in I_t\}
  \quad \text{where }
  \ell_t  = \text{logit}^{-1}(L_{t}), 
  \\
   L_{t} | L_{t-1} &\sim N\big(\phi\, L_{t-1}, \sigma^2_L\big). 
\end{align*}
We set $\theta_0=3.5$, $\sigma_L=1$ and $\phi=0.9$ and assume these values are known. Parameters $\theta_0$ and $\sigma_L$ of the latent AR(1) control the regularity of the baseline intensity. The self-exciting kernel, $K_H$, is of parametric form, namely, we use is the exponential decaying intensity \citep{dassios2013exact}:
\begin{align}
\mathbb{R}^+ \times \mathbb{R}^+ &\rightarrow \mathbb{R}^+ \\
(\nu, \tau) &\mapsto \theta_1 \theta_2 \exp\Big(-\theta_2 (\tau - \nu)\Big).
\end{align}
We have constructed the parameterisation so that $\theta_1$ controls the strength of the effect of past data and that $\theta_2$ controls the shape of the relationship with past data. Notice that, ignoring $\theta_1$, $K_H$ is in the form of a probability distribution on $\tau-\nu$.
The baseline $\lambda_b$ and the self-excitement kernel, once combined, give the overall Hawkes process rate at $\tau$ given the event before time $\tau$:
\begin{align} \label{eq:Hawkes}
\lambda(\tau) = \theta_0 \sum_{i=1}^{n_i} \ell_t \mathbf{1}\{\tau\in I_t\} + \theta_1 \sum_{\upsilon_j \leq \tau} \theta_2\exp\Big(-\theta_2 (\tau - \upsilon_j)\Big).
\end{align}

The observed data generated by the Hawkes process with this intensity is a set of ordered events $\nu_1$, $\nu_2$,\ldots of random cardinal. The data can be sliced according to our partition of the real axis into nonoverlapping intervals:
\[
  y_t=\{\nu_j: \nu_j\in I_t\}.
\]
The process $y_t$, with the latent $L_t$ does not form a state space model: given the parameters, the set $y_t$ depends on $L_t$, but also on the whole history $y_1,\ldots y_{t-1}$. Nevertheless, the process $x_t=(L_t, y_{1:t-1})$ forms a Markov chain, and it is clear that the process $y_t$ with the latent $x_t$ forms a state space model.
Given $x_{t-1}$ and $\theta$, the proposal $q_t$ which we use to simulate $x_t$ is as follows: we draw $L_t|L_{t-1}\sim N(\phi L_{t-1},\sigma_L^2)$ and we complete $y_{1:t-2}$ with the observed $y_{t-1}$. Yet, the simulated $\tilde y_t$ are drawn as a realisation of the Hawkes process on $I_t$ given $\theta$ and $x_t=(L_t,y_{1:t-1})$.

Note that, in either case, it is possible that $y_t$ or $\tilde y_t$ is the empty set. This complicates definitions of summary statistics, so we append some imaginary events $\upsilon_{A_t}$ and $\upsilon_{B_t}$: we set $y_t' = \upsilon_{A_t} \cup y_t \cup \upsilon_{B_t}$ and $\tilde y_t'=\upsilon_{A_t} \cup \tilde  y_t \cup \upsilon_{B_t}$, where $\upsilon_{A_t} = 10t - 10$ and $\upsilon_{B_t} = 10t$. The following summary statistics are now defined on $y_t'$ (and likewise on $\tilde y_t'$): the number of events $n_t$ in the interval, the sum of squares of inter-event times, 
\begin{align*}
\text{diff}^2_t := \sum_{j=1}^{n_t - 1} (\upsilon_j - \upsilon_{j-1})^2; 
\end{align*}
the sum of cubes of inter-event times, 
\begin{align*}
\text{diff}^3_t := \sum_{j=1}^{n_i - 1} (\upsilon_j - \upsilon_{j-1})^3;
\end{align*}
and the minimum of inter-event times, 
\begin{align*}
\text{md}_t := \min (\upsilon_j - \upsilon_{j-1}).
\end{align*}

These summary statistics were used to construct estimators for $\theta$, and $L$, based on the summary statistics. To construct the estimators, we performed linear regression adjustment on these summary statistics \citep{beaumont_approximate_2002}. The distance $d$ is the Euclidean distance on these empirical estimators. Tuning parameters used in the SMC$^2$ procedure were the following $N_{\theta} = 10^5, N_x = 20, N_y = 1$. The transition kernel $K_t$ (Algorithm \ref{algo:ABCSMC2update}) is defined in the following manner: 
\begin{align}
\log(\tilde{\theta}) &\sim N \left[ \log(\theta), c \hat{\Sigma_{\theta}} \right], 
\end{align}
where $\hat{\Sigma}$ is the maximum likelihood estimate of the covariance matrix of $\log(\theta^m)$ and $c$ is a fixed tuning parameter set to 0.1. 
Based on our definition of $K_t$, the ratio of transition kernels is derived
\begin{align*}
\frac{K_t(\breve{\theta}^m|\theta^m)}{K_t(\theta^m|\breve{\theta}^m)} = \frac{\breve{\theta_1} \breve{\theta_2}}{\theta_1 \theta_2}.
\end{align*}

\subsubsection{Results}
The synthetic dataset we consider as observed data is simulated with known parameter values $\theta_1~=~\theta_2~=~0.5$. The observations $y_t$ are simulated up to $t = 60$. The true state is shown, in red, in Figure \ref{fig:Hawkes_state}; along with the Bayesian prediction intervals for the marginalised filtering distributions. The results show very good correspondence with the true state, considering the complexity and dependencies within the Hawkes simulation, see Equation~\eqref{eq:Hawkes}. 

The priors were $U(0.3, 0.7)$ for both parameters. The posteriors are shown in Figure \ref{fig:Hawkes_parameter}. The precision of the state inference is not reflected in the parameter inference, parameters. The posterior variance is slightly higher for $\theta_2$, which may be related to how the distance was defined, with respect to estimators on $x_t$ and $\theta_1$. 

This example illustrates how a complex problem, with a nonparametric prior on the functional parameters of a stochastic process, can be included in a state space model. The inference problem is thus more challenging than the parametric problem studied by \citet{rasmussen2013bayesian}. Despite the approximation due to an ABC approach, the results we obtained are interesting. They may deserve more numerical studies to compare carefully there accuracy with methods of proven accuracy such as \citet{reynaud2013inference}.

\section{Discussion and Conclusion} \label{sec:ABCSMC2_discussion}

We have developed a CPS inference method (Algorithm \ref{algo:ABCSMC2update}) for likelihood-free problems without the need for any manual calibration of time-indexed tuning parameters. This algorithm, in some sense, resembles that from Chapter 4, where SABC transitions a set of parameter proposals through a sequence of decreasing and automatically calibrated thresholds. This is one example of the link between sequential algorithms for fixed parameters and particle filters for state parameters. In this chapter, we have constructed an algorithm which performs likelihood-free CPS inference with automatically tuned thresholds. The ABC thresholds $\epsilon_{1:t}$ are automatically calibrated as the algorithm progresses along the time index. It is important to store $\epsilon_{1:t}$, as they are reused within the rejuvenation step to ensure that the resampling-move step preserves the intended approximate posterior distribution. We successfully retrieved the state trajectory, to a high level of precision, in all three examples and also retrieved the static parameters. Inference for the static parameters proved to be more difficult than the state parameters in all three examples. 


As is the case for the SMC$^2$ algorithm \citep{chopin2013smc2}, the computation time scales superlinearly with time $t$. Recently, \citet{crisan2018nested} developed a CPS algorithm which scales linearly with $t$, with the disadvantage that the algorithm is no longer consistent for fixed $N_x$. The memory requirements also became very high. For instance, the skew normal example, had $N_{\theta} = 2^3$, $N_x = 2^3$, and 20 time points. With 16 processors, the computation time was 37 minutes, with 14 GB of memory usage. More efficient data structures for storing full paths are discussed by \citet{jacob2015path}.

Indexing of the variables within Algorithm \ref{algo:ABCSMC2update} became very involved. This is because the algorithms are naturally written as object orientated programs, where the particle filter and SMC$^2$ update communicate with one another while storing their evaluation environment. However, we wrote this algorithm as a functional program since this is the form most interpretable to statisticians.


\section*{Acknowledgements}

Computing facilities were provided by the High Performance Computing (HPC), Queensland University of Technology. This work was supported by the ARC Centre of Excellence for Mathematical and Statistical Frontiers (ACEMS). This work was funded through the ARC Linkage Grant “Improving the Productivity and Efficiency of Australian Airports" (LP140100282) and the ARC Laureate Fellowship Program. At the beginning of this work, Kerrie Mengersen held the Jean-Morlet Chair at CIRM, Aix-Marseille University. The three first authors (Ebert, Pudlo, Mengersen) took advantage of the Chair's invitation program to begin this research work. Christopher Drovandi was supported by an Australian Research Council Discovery Project (DP200102101).

\vskip 0.2in
\bibliography{ABCSMC2}

\appendix
\section{Proof of Proposition~\ref{pro:expected_value}}
\label{sub:proof_expected_value}
The proof is done by induction on $t=1,2,\ldots$.

\paragraph{Base case: $t=1$.}   Since
\[
  Z_1^mW_1^{m,n}=N_x^{-1}w_1^{m,n}= \frac{\ds p\Big( x_1^{m,n} \Big|
    \theta^m\Big) }{\ds N_x
    q_t\Big( x_1^{m,n} \Big| \theta^m\Big)
  }N_y^{-1}\sum_{i=1}^{N_y} \mathbf 1 \{
  d_1^{m,n}(i) \le \epsilon\},
\]
we have
\begin{align}
  \check\varphi_1
  & = N_\theta^{-1}\sum_{m=1}^{N_\theta}  N_x^{-1}
    \sum_{n=1}^{N_x} N_y^{-1}\sum_{i=1}^{N_y} \check\varphi_1^{m,n,i}
    \quad \text{where} \label{eq:varphi1mni}
  \\
  \check\varphi_1^{m,n,i}
  & :=  \frac{\ds p\Big( x_1^{m,n} \Big|  \theta^m\Big) }{\ds 
    q_t\Big( x_1^{m,n} \Big| \theta^m\Big) }
    \mathbf 1 \{  d_1^{m,n}(i) \le \epsilon_1\} \,
    \varphi(x_1^{m,n},\theta^m). \notag
\end{align}
Because of the sampling distributions of the algorithm:
\[
  \theta^m, x_1^{m,n}, \tilde y_1^{m,n}(i) \sim p(\theta)
  q_1(x_1|\theta) p(\tilde y_1| x_1, \theta),
\]
we have
\begin{align}
  \esp(\check\varphi_1^{m,n,i})
  &= \iiint d\theta dx_1 d\tilde y_1
    p(\theta) p(x_1|\theta) p(\tilde y_1|x_1, \theta)
    \mathbf 1 \{  d_1^{m,n}(i) \le \epsilon_1\}
    \varphi(x_1,\theta) \notag
  \\
  &= p_{\epsilon_1}(y_1) \iint d\theta dx_1 p_{\epsilon_1}(x_1,
    \theta|y_1) \varphi(x_1, \theta) \label{eq:varphi1mni_expected}
\end{align}
by the definition \eqref{eq:target_1} of the target at $t=1$. Since
this last right hand side does not depend on $m,n$ and $i$, 
by plug in this result into \eqref{eq:varphi1mni}, we get the
desired result for $t=1$.

\subsubsection{Inductive step:} Let $\mathscr F_{t-1}$ is the
$\sigma$-field generated by all random draws of the ABC-SMC$^2$
algorithm before entering the $t$-th update. As above, since
$Z_t^mW_t^{m,n} = N_x^{-1}Z_{t-1}^m w_t^{m,n}$, we have
\begin{align}
  \check\varphi_t
  & = N_\theta^{-1}\sum_{m=1}^{N_\theta} Z_{t-1}^m \, N_x^{-1}
    \sum_{n=1}^{N_x} N_y^{-1}\sum_{i=1}^{N_y} \check\varphi_t^{m,n,i}
    \quad \text{where} \label{eq:varphitmni}
  \\
  \check\varphi_t^{m,n,i}
  & :=  \frac{\ds p\Big( x_t^{m,n} \Big|  \theta^m, x_{t-1}^{m,a^{m,n}_t}\Big) }{\ds 
    q_t\Big( x_t^{m,n} \Big| \theta^m, x_{t-1}^{m,a^{m,n}_t}\Big) }
    \mathbf 1 \{  d_t^{m,n}(i) \le \epsilon_t\} \,
    \varphi(x_t^{m,n},\theta^m). \notag
\end{align}
And, during the $t$-th update, we draw $a_t^{m,n}|\mathscr F_{t-1}\sim \mathcal
M\Big(W_{t-1}^{m,a}, a=1,...n\Big)$, and
\[
  x_t^{m,n}, \tilde y_t^{m,n}(i) \Big|\mathscr F_{t-1}, a_t^{m,n}=a
  \,\sim
  p(x_t|x_{t-1}^{m,a}, \theta)   p(\tilde y_t |x_t, \theta).
\]
Thus,
\begin{align*}
  \esp\Big( \check\varphi_t^{m,n,i} \Big| \mathscr F_{t-1}\Big)
  &= \sum_{a=1}^{N_x}W_{t-1}^{m,a}\int dx_t d\tilde y_t
    p(x_t|\theta^m, x_{t-1}^{m,a})   p(\tilde y_t | x_t, \theta^m)
    \mathbf 1\{ d(\tilde y_t, y_t \le \epsilon_t \}
    \varphi(x_t, \theta^m).
\end{align*}
Thus,
\begin{gather*}
  \esp\Big( \check\varphi_t\Big| \mathscr F_{t-1}\Big) =
  N_{\theta}^{-1}\sum_{m=1}^{N_x} Z_{t-1}^m\sum_{n=1}^{N_x}
  W_{t-1}^{m,n} \,
  \psi(x_{t-1}^{m,n},\theta^m) \quad \text{where, for all }\theta
  \text{ and } x_{t-1},
  \\
  \psi(x_{t-1},\theta) := \int dx_t d\tilde y_t
  p(x_t|\theta, x_{t-1})   p(\tilde y_t | x_t, \theta)
  \mathbf 1\{ d(\tilde y_t, y_t \le \epsilon_t \}
  \varphi(x_t, \theta^m).
\end{gather*}
Using the induction hypothesis, we get
\[
  \esp(\check\varphi_t) = p_{\epsilon_{1:t-1}}(y_{1:t-1}) \iint
  d\theta dx_t p_{\epsilon_{1:t-1}}(x_{t-1}, \theta|y_{1:t-1}) \psi(x_{t-1},\theta)
\]
This gives the desired results because of the inductive definition
of the target given in \eqref{eq:target_t}. 

\section{Proof of Propostion~\ref{pro:LLN}}
\label{sub:proof_LLN}
We start by proving the consistency at $t=1$.
For the moment, fix $n$ and $i$ and consider the
$\check\varphi_1^{m,n,i}$, $m=1,\ldots, N_\theta$, introduced in
Equation~\eqref{eq:varphi1mni}.
These $N_\theta$ random variables are iid and
\[
  \esp(\check\varphi_1^{m,n,i})=p_{\epsilon_1}(y_1)
  \iint d\theta dx_1 p_{\epsilon_1}(x_1,\theta|y_1)
  \varphi(x_1, \theta)
\]
because of Equation~\eqref{eq:varphi1mni_expected}.
By the strong law of large number, we obtain that
\[
  \frac{1}{N_\theta} \sum_{m=1}^{N_\theta} \check\varphi_1^{m,n,i}
  \to p_{\epsilon_1}(y_1)
  \iint d\theta dx_1 p_{\epsilon_1}(x_1,\theta|y_1)
  \varphi(x_1, \theta) \quad \text{a.s.}
\]
To obtain the desired result at $t=1$, it is enough to average the
above result over $n=1,\ldots, N_x$ and $i=1,\ldots, N_y$.

\bigskip

We prove now the consitency at $t>1$.
Likewise, because of Equation~\eqref{eq:varphitmni},  it is enough to prove
that, for any fixed $n$ and $i$,
\begin{equation}
  \frac{1}{N_\theta}\sum_{m=1}^{N_\theta} Z_{t-1}^m
  \check\varphi_t^{m,n,i} \to
   \iint d\theta
    \, dx_t \,
    p_{\epsilon_{1:t}}(x_t, \theta|y_{1:t})  \, \varphi(x_t,\theta)
    \quad \text{a.s.}
    \label{eq:const}
  \end{equation}
Let $\mathscr G_m$ be the $\sigma$-field generated by $\theta^m$,
$a_s^{m,n}$, $x_s^{m,n}$, $y_s^{m,n}(i)$ for $s=1,\ldots, t$,
$n=1,\ldots, N_x$ and $i=1,\ldots, N_y$.
Since we assume that we never resample the $\theta^m$'s, these
$\sigma$-fields are independant.
Now, fix $n$ and $i$.
The random variables $Z_{t-1}^m\check\varphi_t^{m,n,i}$ are $\mathscr
G_m$-mesurable, thus they are independant. And it is clear that they
are identically distributed. The strong law of large numbers concludes
the proof.
\end{document}